\documentclass{article}
\usepackage{geometry}
\usepackage[utf8]{inputenc}
\usepackage{hyperref}
\usepackage{tikz}
\usetikzlibrary{shadows}
\usepackage{amsmath}  
\usepackage{booktabs}
\usepackage{newunicodechar}
\newunicodechar{ }{\,}
\newunicodechar{ }{\,}
\newunicodechar{≈}{\approx}
\newunicodechar{≈}{\approx}
\newunicodechar{∼}{\textasciitilde}
\newunicodechar{−}{\textminus}
\usepackage{booktabs}
\usepackage{pgfplotstable}
\pgfplotsset{compat=1.17}

\usetikzlibrary{positioning, shapes.geometric}

\usepackage{listings}
\title{How Cybersecurity Behaviors affect the Success of Darknet Drug Vendors: A Quantitative Analysis}
\author{\small Syon Balakrishnan and Aaron Grinberg}
\date{\small July 2025}

\begin{document}

\maketitle
\begin{abstract}
Understanding behavioral drivers of success in illicit digital marketplaces is critical for developing effective enforcement strategies and understanding digital commerce evolution, as darknet drug markets represent a growing share of the total drug economy. This study employs quantitative regression analysis of 50,000+ listings from 2,653 vendors in the Agora marketplace (2014-2015), examining relationships between cybersecurity signaling (PGP encryption mentions), product diversification, and commercial success through nested regression specifications controlling for reputation, pricing, and category-specific factors. Product diversification emerges as the dominant predictor of vendor scale, increasing the odds of large vendor status by 169\% per additional category, while PGP encryption signaling functions primarily as a professional marker rather than an independent success factor. Vendor success depends on portfolio breadth rather than specialization, with category-specific enforcement creating differential market constraints. Successful vendors operate as diversified enterprises capable of rapid pivoting between product categories, requiring targeted enforcement towards diversified vendors based on coordinated multi-category enforcement approaches rather than traditional substance-specific targeting strategies.
\end{abstract}
\section{Introduction}

\subsection{Background and Context}
The darknet represents a collection of encrypted networks accessible only through specialized software like Tor (The Onion Router), creating anonymous digital marketplaces that operate beyond conventional Internet infrastructure and law enforcement surveillance capabilities. These platforms have fundamentally transformed illicit commerce by enabling global drug distribution networks that bypass traditional territorial controls, physical supply chains, and face-to-face transaction requirements. Notable examples include Silk Road (2011-2013), which facilitated over \$1.2 billion in transactions before its closure, AlphaBay (2014-2017) with over 300,000 users, and Agora (2013-2015) which became the largest marketplace following Silk Road's shutdown, demonstrating the persistent demand for anonymous commercial platforms~\cite{christin2013, martin2014}. It's estimated that the darknet drug markets represent \$300,000+ of the total drug market ~\cite{soska2015measuring} and has been shown to be growing over time. ~\cite{chainalysis2023darknet}

\subsection{Research Motivation}
The economic significance of darknet drug markets has grown substantially, with current estimates suggesting annual revenues exceed \$300 million globally and involving hundreds of thousands of participants across international borders~\cite{unodc2020}. Unlike traditional drug markets characterized by territorial control, social embeddedness, and physical enforcement vulnerabilities, darknet markets operate through sophisticated technical infrastructure that creates new challenges for law enforcement while enabling novel forms of criminal enterprise. Understanding the behavioral factors that drive vendor success in these anonymous environments has become critical for developing effective intervention strategies, as traditional enforcement approaches prove insufficient against digitally-native criminal operations that leverage advanced cybersecurity practices for operational security.

\subsection{Literature Gaps}
Despite growing policy interest, several critical knowledge gaps limit our understanding of darknet market dynamics. \textbf{First}, existing research emphasizes descriptive market analysis and user behavior studies while providing limited quantitative analysis of vendor success factors and business strategies. \textbf{Second}, the relationship between cybersecurity practices and commercial outcomes remains underexplored, with most studies treating encryption usage as a binary security measure rather than examining its role as a strategic signaling mechanism. \textbf{Third}, product diversification strategies and their impact on vendor survival and scale have received minimal empirical attention, despite theoretical suggestions that portfolio breadth may provide insurance against enforcement and demand volatility.

\subsection{Research Questions}
This study addresses three fundamental research questions: (1) How do cybersecurity signaling behaviors, specifically PGP encryption mentions, relate to vendor commercial success in anonymous digital marketplaces? (2) What role does product diversification play in vendor scaling and competitive positioning relative to other strategic factors? (3) How do category-specific market conditions and enforcement patterns influence vendor success across different drug types?

\subsection{Methodology Overview}
Our analysis employs nested regression modeling using a large-scale dataset of over 50,000 listings from 2,653 vendors operating in the Agora marketplace between 2014 and 2015. We implement progressive model specifications that isolate encryption signaling effects while controlling for reputation accumulation, product diversification, pricing strategies, and category-specific market conditions.

\subsection{Key Contributions}
This research makes several important contributions to both theory and practice. \textbf{First}, we provide the first large-scale quantitative analysis of cybersecurity signaling effects on vendor performance, revealing that encryption mentions function primarily as professional signals rather than independent success drivers. \textbf{Second}, we demonstrate that product diversification represents the dominant pathway to vendor success, fundamentally challenging assumptions about specialization advantages in illicit markets. \textbf{Third}, we identify category-specific constraints that create systematic differences in vendor scaling potential, with important implications for enforcement resource allocation and market dynamics.

\subsection{Paper Structure}
The remainder of this paper is organized as follows. Section 2 reviews relevant literature on darknet markets, digital trust mechanisms, and illicit commerce strategies. Section 3 describes our data sources, variable construction, and analytical methodology. Section 4 presents empirical results across multiple model specifications. Section 5 discusses theoretical and practical implications, while Section 6 concludes with policy recommendations and future research directions.

\section{Literature Review}

\subsection{Darknet Markets in Digital Commerce Ecosystems}

The darknet represents an encrypted network operating beyond conventional Internet infrastructure and law enforcement surveillance, fundamentally transforming illicit commerce by creating new channels for drug distribution that bypass traditional territorial controls and enforcement mechanisms~\cite{christin2013, martin2014}. These digital marketplaces have evolved from simple bulletin boards to sophisticated e-commerce platforms featuring vendor reputation systems, escrow services, and customer feedback mechanisms that mirror legitimate online retail environments. Notable examples include Silk Road (2011-2013), which facilitated over \$1.2 billion in transactions before closure, AlphaBay (2014-2017) with over 300,000 users, and Agora (2013-2015), which became the largest marketplace following Silk Road's shutdown~\cite{soska2015measuring, aldridge2016}.

Drug sales represent the dominant use case for these markets, with recent estimates suggesting global revenues exceed \$300 million annually and involving hundreds of thousands of participants across international borders~\cite{unodc2020}. Unlike conventional drug markets characterized by territorial control, social embeddedness, and face-to-face trust relationships, darknet markets mediate all interactions through pseudonymous profiles, customer reviews, and product listings~\cite{vanhout2014}. This fundamental shift from physical to digital commerce replaces direct interpersonal trust with digital reputation systems and behavioral signals, requiring vendors to develop entirely new forms of strategic behavior to establish credibility and attract customers in anonymous environments.

\begin{table}[ht]
\caption{Vendor Success Mechanisms in Darknet Markets and Supporting Literature}
\label{tab:mechanisms}
\centering
\begin{tabular}{|p{3.5cm}|p{4.5cm}|p{4.5cm}|}
\hline
\textbf{Mechanism} & \textbf{Rationale} & \textbf{Representative Literature} \\
\hline
Reputation Signaling & High ratings and feedback scores attract buyer trust and drive repeat transactions. & Christin (2013)~\cite{christin2013}, Janetos \& Tilly (2017)~\cite{janetos2017}, Eschenbaum \& Liebert (2021)~\cite{eschenbaum2021}, Andrei \& Veltri (2025)~\cite{andrei2025}, McBride (2023)~\cite{mcbride2023} \\
\hline
Product Diversification & Vendors listing across multiple categories may reduce volatility and appeal to wider demand segments. & Crépino et al. (2019)~\cite{crepino2019}, Szigeti \& Molnár (2023)~\cite{szigeti2023}, Torre (2018)~\cite{torre2018}, Fonseca dos Reis et al. (2024)~\cite{fonseca2024multihomers}, Aitken et al. (2019)~\cite{aitken2019} \\
\hline
Encryption Signaling & PGP use may signal professionalism or reduce exposure to legal risk in vendor-buyer communication. & Dwyer et al. (2022)~\cite{dwyer2022}, UNODC (2020)~\cite{unodc2020}, Soska \& Christin (2015)~\cite{soska2015measuring} \\
\hline
Price Strategy & Vendors may use competitive or prestige pricing to differentiate and influence buyer perceptions. & Aldridge \& Décary-Hétu (2016)~\cite{aldridge2016}, Zaunseder \& Koenig (2020)~\cite{zaunseder2020}, Barratt \& Aldridge (2016)~\cite{barratt2016} \\
\hline
\end{tabular}
\end{table}

\subsection{Trust and Reputation Mechanisms}

Reputation signaling constitutes the first critical dimension of vendor success, as customer feedback systems aggregated at the listing level serve as substitutes for in-person trust relationships in anonymous digital environments. Vendors must accumulate positive ratings and transaction volumes to demonstrate reliability, with higher-rated sellers attracting more traffic, commanding premium prices, and converting repeat customers more effectively. Past research consistently links high feedback scores to higher sales volumes, price premiums, and longer operational lifecycles, with some studies distinguishing between costly signals such as verified seniority and escrow usage versus superficial signals such as self-promotional descriptions~\cite{janetos2017, eschenbaum2021, andrei2025}. 

McBride~\cite{mcbride2023} characterizes this reliance on digital reputation mechanisms as a form of "yelpification" of illicit commerce, where algorithmic feedback systems directly shape market outcomes and vendor survival. The reputation economy in darknet markets exhibits several unique characteristics compared to legitimate e-commerce platforms. First, the irreversibility of transactions and limited dispute resolution mechanisms place greater weight on ex-ante reputation assessment. Second, the anonymous nature of transactions prevents buyers from developing direct vendor relationships, making aggregated feedback scores the primary trust mechanism. Third, the illegal nature of transactions creates additional reputational dimensions around discretion, shipping reliability, and operational security that do not exist in legitimate commerce~\cite{christin2013}.

\subsection{Product Diversification Strategies}

Product diversification represents the second strategic dimension, as vendors who list products in multiple drug categories can hedge against demand uncertainty, maintain consistent listing presence during supply disruptions, and appeal to broader buyer segments seeking convenient one-stop-shop experiences. Torre~\cite{torre2018} frames this diversification as an intentional managerial strategy analogous to Ansoff's growth matrix, where vendors systematically expand product lines to stabilize revenues and reduce category-specific risks. These risks include differential enforcement targeting, as law enforcement agencies prioritize certain drug categories over others based on perceived threat levels and public health impacts.

The U.S. Drug Enforcement Administration's 2024 National Drug Threat Assessment explicitly identifies fentanyl as "the nation's greatest and most urgent drug threat," with synthetic opioids and stimulants receiving the vast majority of enforcement attention—fentanyl and other synthetic opioids are responsible for approximately 70\% of lives lost, while methamphetamine and other synthetic stimulants are responsible for approximately 30\% of deaths~\cite{dea2024ndta}. This differential prioritization is reflected in prosecution patterns, with fentanyl cases increasing by 246 percent over the last five fiscal years to become the second most common federal drug prosecution, while marijuana cases account for only 3 percent of all federal drug cases~\cite{ussc2024annual}. By maintaining presence across multiple categories, vendors can continue operations even when enforcement efforts intensify around specific drug types.

The strategic importance of diversification becomes particularly evident when considering the inherent volatility of darknet vendor participation. The useful lifetimes of darknet vendors are often remarkably short, with almost 45\% of Silk Road sellers remaining active for less than 30 days, although average tenures range between 90 and 100 days across different platforms~\cite{soska2015measuring}. This high turnover rate underscores the precarious nature of vendor operations and highlights diversification as a potential survival mechanism, with blockchain-based network analyses showing that multihoming strategies—simultaneous presence in multiple marketplaces—support ecosystem resilience during market shutdowns and law enforcement actions~\cite{fonseca2024multihomers}.

\subsection{Cybersecurity and Operational Security}

Cybersecurity signaling and operational security practices constitute the fourth strategic dimension, where vendors advertise encryption capabilities and secure communication channels to signal both technical sophistication and commitment to protecting customer privacy. Secure communication through tools like PGP (Pretty Good Privacy) has become both a cultural and operational norm in darknet commerce, with vendors frequently posting public encryption keys in product listings to enable buyers to encrypt sensitive messages containing delivery addresses and payment information~\cite{dwyer2022}.

Although such practices may provide genuine operational security benefits by reducing traceability and evidence generation, they also function as powerful professionalism signals within darknet communities where technical competence serves as a proxy for overall reliability and trustworthiness. Encryption technologies such as PGP create structural obstacles for law enforcement in darknet drug markets, as these tools hinder monitoring, evidence collection, and intervention capabilities~\cite{unodc2020}. Within vendor communities, encryption functions simultaneously as both a normative expectation and a practical safeguard, with experienced sellers often advocating the use of PGP as a marker of professional competence~\cite{dwyer2022}.

However, despite the widespread advocacy for encryption among experienced sellers documented in forum analyses, empirical research has yet to quantitatively link PGP usage patterns to actual commercial outcomes, representing a significant gap in our understanding of how cybersecurity behaviors translate into market success. The tension between operational security needs and commercial signaling requirements creates complex strategic considerations for vendors operating in technically sophisticated criminal environments.

\subsection{Pricing and Market Positioning}

Vendors in darknet markets face complex pricing decisions that extend beyond simple profit maximization, as pricing strategies must simultaneously signal product quality, operational reliability, and risk tolerance while navigating the unique constraints of anonymous commerce. Vendors may employ competitive pricing to increase transaction volumes and build market share among price-sensitive buyers, or alternatively adopt prestige pricing strategies that signal superior product quality or exclusive access to premium supplies~\cite{zaunseder2020, barratt2016}.

Zaunseder and Koenig~\cite{zaunseder2020} demonstrate that pricing decisions reflect complex interactions between vendor reputation, perceived product quality, and risk tolerance, with these relationships operating through multiple interconnected mechanisms that create sophisticated market dynamics. High-reputation vendors can leverage their established credibility to command premium prices, as buyers are willing to pay more for perceived reliability and quality assurance from trusted sellers, while newer vendors with limited transaction histories must often engage in competitive pricing strategies to build market share and accumulate positive feedback necessary to justify higher prices in future transactions.

Soska and Christin~\cite{soska2015measuring} document sophisticated pricing dynamics where vendors temporarily pause listings to manage customer reviews and reset price expectations, revealing how pricing operates as a dynamic tool for reputation management rather than a static market positioning strategy. These temporal pricing patterns demonstrate vendors' strategic awareness that pricing decisions have both immediate revenue effects and longer-term reputational consequences. Barratt and Aldridge~\cite{barratt2016} emphasize that high prices, when paired with detailed product descriptions and satisfaction guarantees, can serve as trust signals that communicate vendor confidence in product quality and service delivery.

\subsection{Research Gaps and Theoretical Framework}

Despite growing academic and policy interest in darknet markets, several critical knowledge gaps limit our understanding of vendor success factors and market dynamics. Existing research emphasizes descriptive market analysis and user behavior studies while providing limited quantitative analysis of vendor success factors and business strategies. Most studies focus on market size estimation, drug availability patterns, or user demographics rather than examining the specific behavioral factors that drive vendor performance and longevity~\cite{christin2013, aldridge2016}.

The relationship between cybersecurity practices and commercial outcomes remains underexplored, with most studies treating encryption usage as a binary security measure rather than examining its role as a strategic signaling mechanism. Although qualitative studies have identified vendor professionalism typologies and marketing strategies~\cite{vanhout2014, munksgaard2016}, they rarely treat encryption practices as measurable factors that can be empirically linked to performance outcomes. This analytical gap limits our understanding of how vendors navigate the tension between operational security needs and commercial signaling requirements in technically sophisticated criminal environments.

Product diversification strategies and their impact on vendor survival and scale have received minimal empirical attention, despite theoretical suggestions that portfolio breadth may provide insurance against enforcement and demand volatility. The few existing studies focus on single-category analysis or treat diversification as a control variable rather than examining it as a primary strategic mechanism~\cite{crepino2019}.

Additionally, existing research lacks comprehensive frameworks that integrate multiple success mechanisms while controlling for category-specific market conditions and enforcement patterns. Most studies examine individual factors in isolation rather than testing competing explanations for vendor success within unified analytical frameworks.

This study addresses these theoretical and policy gaps by providing the first large-scale quantitative analysis of encryption signaling effects on vendor performance outcomes. Using data from over 2,600 vendors operating in the Agora marketplace between 2014 and 2015, we examine whether vendors mentioning PGP encryption capabilities achieve systematically different outcomes in terms of business scale, customer satisfaction, and competitive positioning. Our analytical approach nests encryption effects within a comprehensive framework that controls reputation accumulation, product diversification, pricing strategies, and category-specific market conditions, allowing us to isolate the unique contribution of cybersecurity signaling to commercial success while accounting for confounding behavioral and structural factors.

\section{Methods}
\label{sec:methods}

Our analysis employs quantitative regression modeling to examine the relationship between cybersecurity behaviors and vendor success in darknet drug markets, utilizing a large-scale dataset from the Agora marketplace and implementing nested regression specifications to isolate causal mechanisms. All data processing and statistical analyses were conducted in Python 3.9, using pandas for data manipulation~\cite{mckinney2010pandas}, NumPy for numerical operations~\cite{harris2020numpy}, and statsmodels for econometric modeling~\cite{seabold2010statsmodels}.

\subsection{Data Sources and Sample Construction}

Our analysis is based on a structured data set of over 100,000 unique darknet listings from Agora, one of the largest drug-focused marketplaces operating between 2014 and 2015. Agora was a darknet marketplace operating in the Tor network that featured an eBay-like interface with vendor listings, reputation systems, and search functionality~\cite{darknetone2022, wikipedia2025agora}. By September 2014, Agora had become the largest darknet market with over 16,000 listings. Its achieved total sales volume of approximately \$220.7 million USD over its operational lifetime~\cite{christin2015darknet, foley2019measuring}.
We selected Agora for several methodological reasons. First, Agora operated during a critical period (2013-2015) when darknet markets were maturing from experimental platforms to sophisticated commercial ecosystems. Second, unlike many contemporaneous markets that experienced sudden law enforcement closures, Agora voluntarily shut down with advance notice, reducing potential selection bias from vendor exodus patterns. Third, the availability of comprehensive listing data through the usheep archive provides unparalleled granularity for quantitative analysis compared to fragmented datasets from other markets. Finally, Agora's position as the largest market following Silk Road's closure makes it representative of mature darknet commerce practices.
The dataset we use was originally posted by a Reddit user known as ``usheep''~\cite{usheep2015}, who extracted the listings from Agora's surface-web front-end through HTML scraping—an automated process for extracting data from website code—and later publicly released the archive. A Kaggle user, \texttt{philipjames11}, subsequently cleaned and structured this data set into tabular format, removing duplicates and averaging prices for listings that appeared multiple times~\cite{philipjames2017}.

The resulting dataset contains detailed metadata for each listing, including the seller's pseudonymous vendor name, hierarchical product classification, item titles, free-text descriptions containing operational details, averaged Bitcoin prices, geographic shipping claims, textual customer feedback ratings, and notes on potential pricing outliers.

Our sampling procedure implements multiple quality control filters to ensure analytical rigor. We first restrict the sample to listings with standardized Bitcoin pricing and numerical ratings matching the pattern \verb!^\d+\.\d+/5$! to ensure homogeneous measurement units. We then limit our analysis to drug related categories to maintain a focus on illicit substance markets. This filtering process yields approximately 50,000 listings from 2,653 unique vendors, forming our analytical sample.\footnote{\url{https://github.com/syoncodes/agora-regression}}

\subsection{Variable Construction and Measurement}

We construct vendor-level measures to capture marketplace performance and strategic behaviors, organizing variables according to their entry into our nested regression framework.

\subsubsection{Drug Category Classification System}

\begin{figure}[htbp]
\centering
\begin{tikzpicture}[
    box/.style={rectangle, draw, fill=blue!20, text width=3cm, text centered, minimum height=1cm},
    arrow/.style={->, thick}
]
    \node[box] (data) {Raw Agora Dataset\\100,000+ listings};
    \node[box, below=1cm of data] (filter) {Quality Filters\\Bitcoin pricing\\Numerical ratings};
    \node[box, below=1cm of filter] (categories) {Drug Classification\\10-category taxonomy};
    \node[box, below=1cm of categories] (vendor) {Vendor-level\\Aggregation\\2,653 vendors};
    \node[box, below=1cm of vendor] (analysis) {Statistical Analysis\\Nested Regression};
    
    \draw[arrow] (data) -- (filter);
    \draw[arrow] (filter) -- (categories);
    \draw[arrow] (categories) -- (vendor);
    \draw[arrow] (vendor) -- (analysis);
\end{tikzpicture}
\caption{Data Processing and Analysis Workflow}
\label{fig:workflow}
\end{figure}
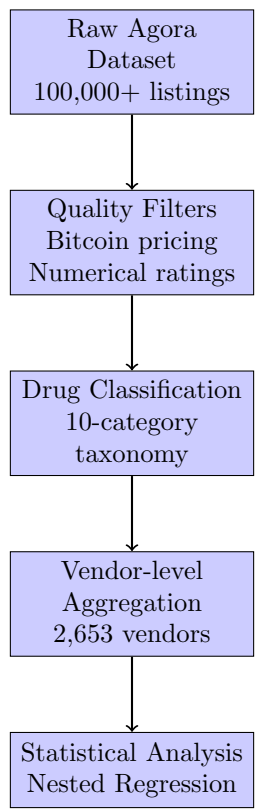

Our analysis employs a ten-category drug taxonomy to control for systematic differences across substance types. Categories include: Cannabis (marijuana and derivatives), Dissociatives (ketamine, PCP), Ecstasy (MDMA variants), Opioids (heroin, prescription opioids), Prescription (benzodiazepines, stimulants), Psychedelics (LSD, psilocybin), RCs (research chemicals), Steroids (performance enhancers), Stimulants (cocaine, amphetamines), and Other/Minor (tobacco, kratom). 

We extract category information from Agora's hierarchical classification system. For listings categorized as ``Drugs/Cannabis/Weed,'' we assign Cannabis as the primary category. This allows us to control for category-specific factors like enforcement intensity and market dynamics. Full classification details appear in Appendix B.

\subsubsection{Dependent Variables}

\textbf{Scale measure}: \texttt{total\_listings} counts unique product listings per vendor, serving as our primary measure of business scope and market presence.

\textbf{Diversification measure}: \texttt{number\_categories} counts distinct drug categories per vendor, capturing strategic diversification that may provide insurance against demand shocks or enforcement.

\textbf{Price positioning}: \texttt{avg\_price\_btc} represents mean Bitcoin price across all vendor listings, normalized within each drug category to account for systematic price differences between substance types. For example, a vendor's cannabis prices are compared to other cannabis prices, not to cocaine prices.

\subsubsection{Independent Variables (By Model Block)}

\textbf{Block 1 - Encryption signaling}: \texttt{pgp\_present} is a binary indicator identifying vendors who mention PGP encryption in any listing description, detected through case-insensitive string matching.

\textbf{Block 2 - Product diversification}: \texttt{number\_categories} enters as a control variable to test whether encryption effects operate independently of diversification strategies.

\textbf{Block 3 - Market positioning}: \texttt{avg\_price\_btc} controls for product quality and market segment effects that may correlate with both encryption use and success.

\textbf{Block 4 - Professional signaling}: \texttt{pct\_professional} measures the percentage of listings containing professional terminology. The variable is calculated as:

\begin{equation}
\text{pct\_professional}_j = \frac{1}{N_j} \sum_{i=1}^{N_j} \mathbf{1}[\text{professional\_terms} \in \text{description}_{ij}]
\end{equation}

where $N_j$ represents total listings for vendor $j$. We identify twenty terms spanning operational security (vacuum sealed, discreet, sealed), quality assurance (lab-tested, purity, pharmaceutical name, blister pack), and customer service (tracking, tracking number, refund, guarantee, reship, business days). Technical proficiency is marked by terms such as escrow and public key. Community engagement is indicated by read profile and erowid references. This tests whether PGP mentions reflect broader professionalism rather than specific encryption effects.

\textbf{Block 5 - Category controls}: Binary indicators for each drug category control for systematic differences in market conditions and enforcement across substance types.

\subsubsection{Success Outcomes}

We define binary indicators capturing different performance thresholds. \texttt{top\_vendor} flags vendors in the top decile by listing count, representing elite performers, while \texttt{large\_vendor} flags vendors above median listing count, representing above-average performers. These allow examination of whether behaviors predict basic viability versus exceptional success.

\subsection{Statistical Modeling Strategy}

\subsubsection{Nested Regression Framework}

Our analysis employs a five-block nested regression structure that progressively adds controls while observing how the encryption effect evolves:

\textbf{Block 1} establishes the baseline PGP-outcome association without controls, providing the raw correlation between encryption signaling and vendor success.

\textbf{Block 2} adds product diversification (\texttt{number\_categories}) to test whether encryption effects merely proxy for underlying business sophistication reflected in category expansion.

\textbf{Block 3} incorporates normalized pricing (\texttt{avg\_price\_btc}) to control for quality positioning and market segmentation that could confound signaling-success relationships.

\textbf{Block 4} includes professional signaling percentage (\texttt{pct\_professional}) to isolate whether PGP represents a distinct mechanism or simply reflects general professionalism.

\textbf{Block 5} adds comprehensive category fixed effects, accounting for substance-specific market conditions, enforcement patterns, and operational requirements.

This progressive specification reveals whether PGP effects attenuate as alternative mechanisms are controlled. Substantial attenuation would suggest encryption signaling proxies for unobserved vendor characteristics rather than representing an independent success factor.

\subsubsection{Model Specifications and Estimation}

We estimate ordinary least-squares models for continuous outcomes and logistic regression models for binary success indicators. All models employ robust standard errors to address heteroskedasticity. Fixed effects for drug categories capture category-specific factors that can influence vendor scale or success regardless of individual vendor strategies. 

All models are estimated using Python 3.9 with the statsmodels library, ensuring reproducible results and robust numerical optimization. Convergence criteria are set to standard tolerances, with alternative optimization algorithms employed when necessary to ensure stable parameter estimates across all model specifications.

\subsubsection{Model Selection and Diagnostic Assessment}

We compare our primary OLS specification against Poisson generalized linear models to account for the count nature of the listing outcome variable. Table~\ref{tab:robust_models} presents the results of the model comparison based on information criteria.

\begin{table}[ht]
\centering
\begin{tabular}{lcc}
\toprule
Model & AIC & BIC\\
\midrule
OLS (raw counts, HC3 SEs) & 27\,385 & 27\,473\\
Poisson (raw count)       & 84\,531 & 51\,297\\
\bottomrule
\end{tabular}
\caption{Comparison of OLS (raw counts) and Poisson models}
\label{tab:robust_models}
\end{table}

The OLS specification demonstrates a superior fit based on model diagnostics and interpretability considerations. Our primary OLS model (Model 5) achieves an AIC of 27,385, with acceptable residual diagnostics, as shown in the diagnostic plots below. This substantial difference in model fit, combined with the interpretability advantages of OLS coefficients, leads us to retain ordinary least squares as our primary specification.

\begin{figure}[ht]
\centering
\includegraphics[width=.6\textwidth]{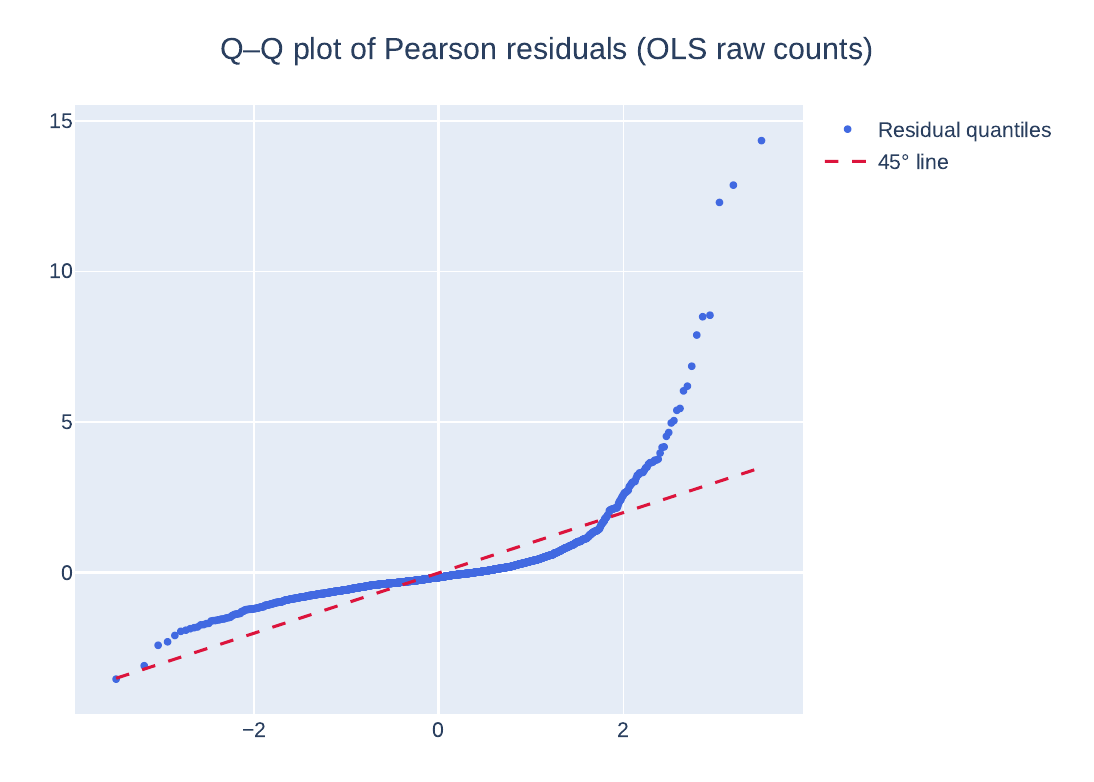}
\caption{Q–Q plot of Pearson residuals from the OLS (raw counts) model}
\label{fig:qq_logols}
\end{figure}

\paragraph{Residual Diagnostics.} Figure~\ref{fig:qq_logols} presents the Q-Q plot of Pearson residuals from our primary OLS specification, demonstrating acceptable adherence to normality assumptions despite some deviation in the extreme tails. This pattern is consistent with the count nature of our outcome variable and the presence of high-volume outlier vendors in our sample.

\begin{figure}[ht]
\centering
\includegraphics[width=.6\textwidth]{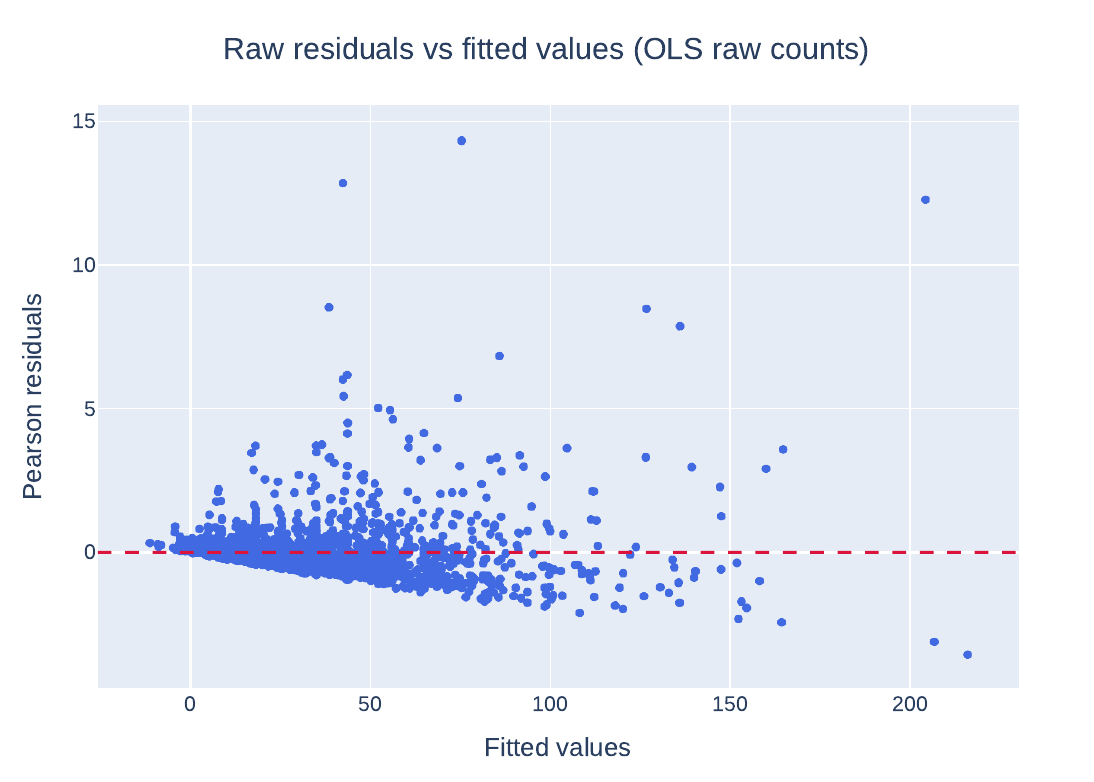}
\caption{Raw residuals vs.\ fitted values for the OLS (raw counts) model}
\label{fig:resid_fitted}
\end{figure}

Figure~\ref{fig:resid_fitted} shows the relationship between the raw residuals and the fitted values, revealing some heteroskedasticity that justifies our use of robust standard errors. The residual pattern shows an increase in variance at higher fitted values, reflecting the greater variability in business models among large-scale vendors compared to smaller operators.

These diagnostic plots confirm that while perfect normality and homoscedasticity are not achieved, the deviations are manageable through robust standard error correction and do not substantially compromise the validity of our statistical inferences.

\section{Results}\label{sec:results}

Our analysis includes 2,653 unique vendors operating in the Agora marketplace. Table~\ref{tab:descriptives} presents descriptive statistics for key variables. The typical vendor maintains 18 listings (median), while the mean vendor maintains 32.7 listings. Vendors diversify across a median of 2 drug categories with a mean of 3.3 categories. The most prolific vendor advertised 815 products across 24 categories. Average customer ratings are 4.83 out of 5 with a median of 4.98. Bitcoin prices range from near zero to 10,892.16 BTC with a median of 0.77 BTC and mean of 16.88 BTC. Of the 2,653 vendors, 139 (5.2\%) mention PGP encryption in their listings.

\begin{table}[htbp]
  \centering
  \scriptsize
  \setlength\tabcolsep{4pt}
  \begin{tabular}{lrrrrr}
    \toprule
    Variable              &   Mean & Median &    SD &    Min &      Max \\
    \midrule
    total\_listings       &  32.68 & 18.00 & 49.25 &  1.00 &  815.00 \\
    num\_categories       &   3.33 &  2.00 &  2.80 &  1.00 &   24.00 \\
    avg\_price\_btc       &  16.88 &  0.77 &260.10 &  0.00 &10892.16 \\
    avg\_rating           &   4.83 &  4.98 &  0.55 &  0.00 &    5.00 \\
    \bottomrule
  \end{tabular}
  \caption{Descriptive statistics ($N=2\,653$ vendors).}
  \label{tab:descriptives}
\end{table}

\begin{figure}[htbp]
\centering
\includegraphics[width=.85\textwidth]{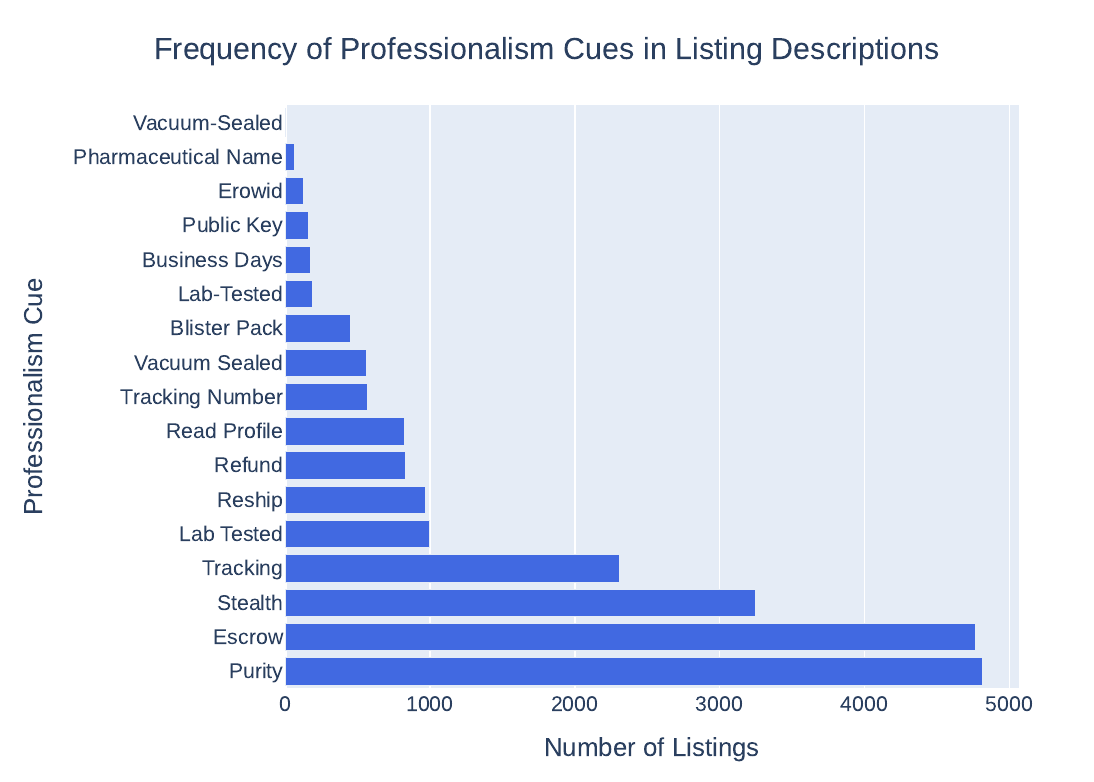}
\caption{Mean share of listings per vendor that include any professionalism cue}
\label{fig:bar_prof_cues}
\end{figure}

Professional signaling patterns differ markedly between vendor types. Figure~\ref{fig:bar_prof_cues} shows that PGP vendors use professional terminology at higher rates than non-PGP vendors, with over 50\% of PGP vendor listings containing professionalism cues compared to approximately 35\% for non-PGP vendors. This suggests that encryption signaling may be part of a broader professionalization strategy rather than an isolated security practice.

\subsection{Vendor Scale Analysis}

\begin{figure}[htbp]
\centering
\includegraphics[width=.75\textwidth]{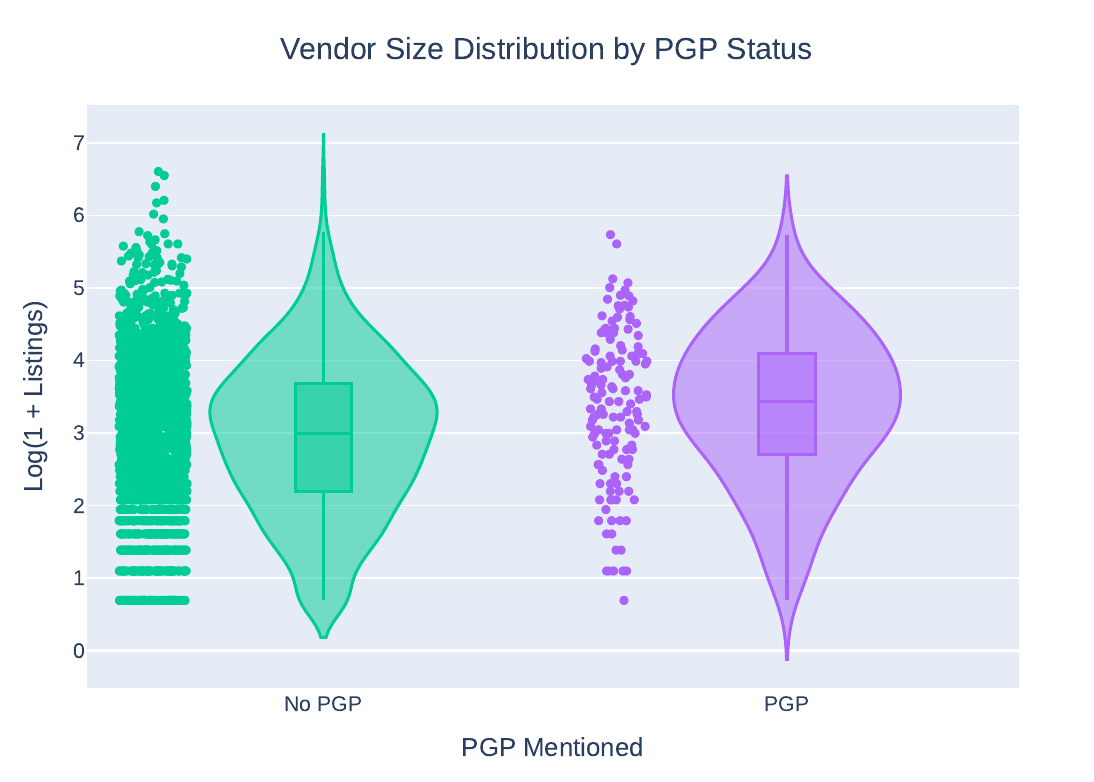}
\caption{Distribution of total listings (log scale) by PGP mention}
\label{fig:box_pgp}
\end{figure}

Figure~\ref{fig:box_pgp} compares the distribution of listings between vendors who mention PGP encryption and those who do not. PGP vendors consistently operate at larger scales across the distribution, with median PGP vendors maintaining approximately twice as many listings as median non-PGP vendors.

Table~\ref{tab:ols_size_nested} shows how PGP and diversification effects evolve across model specifications. In the bivariate specification (Model 1), PGP presence creates a statistically significant effect of 11.578 additional listings ($p < .01$). When product diversification is added in Model 2, the PGP coefficient becomes not statistically significant while the number of categories shows a significant positive association (8.674 additional listings per category, $p < .001$). This pattern remains stable across all subsequent models.

\begin{table}[htbp]
  \centering
  \scriptsize
  \setlength\tabcolsep{4pt}
  \begin{tabular}{lrrrrr}
    \toprule
    & \multicolumn{5}{c}{\textbf{DV: Total listings}}\\
    \cmidrule(lr){2-6}
    Variable & Model 1 & Model 2 & Model 3 & Model 4 & Model 5\\
    \midrule
    Intercept               & 32.076*** & 3.531     & 3.443     & -17.610*** & -14.730***\\
    PGP present             & 11.578**  & 4.927     & 4.984     & 4.676      & 5.389\\
    \# Categories           &           & 8.674***  & 8.675***  & 8.645***   & 11.810***\\
    Avg.\ Price (BTC)       &           &           & 0.005     & 0.005      & 0.005\\
    Pct.\ Professional      &           &           &           & 0.931      & 2.827\\
    Avg.\ Rating            &           &           &           & 4.342***   & 4.106***\\
    \addlinespace
    \multicolumn{6}{l}{\emph{Primary drug class}}\\
    Cannabis                &           &           &           &            & 0.714\\
    Dissociatives           &           &           &           &            & -3.300*\\
    Ecstasy                 &           &           &           &            & -2.377*\\
    Opioids                 &           &           &           &            & -8.051***\\
    Other Minor             &           &           &           &            & 0.000\\
    Prescription            &           &           &           &            & -3.803\\
    Psychedelics            &           &           &           &            & -3.899***\\
    RCs                     &           &           &           &            & -1.150\\
    Steroids                &           &           &           &            & 2.537\\
    Stimulants              &           &           &           &            & -3.834**\\
    \addlinespace
    \multicolumn{6}{l}{\emph{Secondary drug class}}\\
    Cannabis (sec)          &           &           &           &            & 0.714\\
    Dissociatives (sec)     &           &           &           &            & -3.300*\\
    Ecstasy (sec)           &           &           &           &            & -2.377*\\
    Opioids (sec)           &           &           &           &            & -8.051***\\
    Other Minor (sec)       &           &           &           &            & 0.000\\
    Prescription (sec)      &           &           &           &            & -6.402\\
    Psychedelics (sec)      &           &           &           &            & -3.899***\\
    RCs (sec)               &           &           &           &            & -1.150\\
    Steroids (sec)          &           &           &           &            & 2.537\\
    Stimulants (sec)        &           &           &           &            & -3.834**\\
    \midrule
    \multicolumn{6}{l}{\footnotesize Stars: $^{*}p<0.05$, $^{**}p<0.01$, $^{***}p<0.001$. Robust SE: HC3.}\\
    \bottomrule
  \end{tabular}
  \caption{Nested OLS models predicting total listings (Models 1–5)}
  \label{tab:ols_size_nested}
\end{table}

Several drug categories show significant negative associations with vendor scale. Opioid vendors average 8.051 fewer listings ($p < .001$), psychedelic vendors average 3.899 fewer listings ($p < .01$), and stimulant vendors average 3.834 fewer listings ($p < .05$).

\subsection{Customer Rating Analysis}

Table~\ref{tab:ols_rating_full} presents models predicting average customer ratings. PGP presence shows no statistically significant association with ratings ($p = .052$). The number of categories shows a small positive association (0.0129, $p < .05$). The model explains minimal variance in customer satisfaction ($R^2 = .006$).

\begin{table}[htbp]
\centering
\scriptsize
\setlength\tabcolsep{4pt}
\begin{tabular}{lrrrrr}
\toprule
& \multicolumn{5}{c}{\textbf{DV: Average rating}}\\
\cmidrule(lr){2-6}
Variable & Model 1 & Model 2 & Model 3 & Model 4 & Model 5\\
\midrule
Intercept               & 4.830*** & 4.789*** & 4.789*** & 4.795*** & 4.820***\\
PGP present             & 0.067    & 0.043    & 0.043    & 0.044    & 0.043\\
\# Categories           &          & 0.013*   & 0.013*   & 0.013*   & 0.013*\\
Avg.\ Price (BTC)       &          &          & 0.000    & 0.000    & 0.000\\
Pct.\ Professional      &          &          &          & -0.009   & -0.009\\
\addlinespace
\multicolumn{6}{l}{\emph{Primary drug class}}\\
Cannabis                &          &          &          &          & 0.000\\
Dissociatives           &          &          &          &          & -0.023\\
Ecstasy                 &          &          &          &          & -0.019\\
Opioids                 &          &          &          &          & -0.041\\
Other Minor             &          &          &          &          & 0.000\\
Prescription            &          &          &          &          & 0.104***\\
Psychedelics            &          &          &          &          & -0.020\\
RCs                     &          &          &          &          & 0.009\\
Steroids                &          &          &          &          & 0.023\\
Stimulants              &          &          &          &          & -0.016\\
\addlinespace
\multicolumn{6}{l}{\emph{Secondary drug class}}\\
Cannabis (sec)          &          &          &          &          & 0.000\\
Dissociatives (sec)     &          &          &          &          & -0.023\\
Ecstasy (sec)           &          &          &          &          & -0.019\\
Opioids (sec)           &          &          &          &          & -0.041\\
Other Minor (sec)       &          &          &          &          & 0.000\\
Prescription (sec)      &          &          &          &          & -0.127***\\
Psychedelics (sec)      &          &          &          &          & -0.020\\
RCs (sec)               &          &          &          &          & 0.009\\
Steroids (sec)          &          &          &          &          & 0.023\\
Stimulants (sec)        &          &          &          &          & -0.016\\
\midrule
$R^2$                   & 0.001    & 0.003    & 0.003    & 0.003    & 0.006\\
\bottomrule
\multicolumn{6}{l}{\footnotesize $^{*}p<0.05$, $^{**}p<0.01$, $^{***}p<0.001$. $N=2\,653$. Robust SE: HC3.}\\
\end{tabular}
\caption{Nested OLS models predicting average rating (Models 1–5)}
\label{tab:ols_rating_full}
\end{table}
Among drug categories, primary prescription vendors show ratings 0.104 points higher ($p < .001$), while secondary prescription vendors show ratings 0.127 points lower ($p < .001$).

\subsection{Binary Success Outcomes}

Tables~\ref{tab:logit_top_nested} and \ref{tab:logit_large_nested} present nested logistic regression results for achieving top-decile and above-median vendor status. For top-vendor status (Table~\ref{tab:logit_top_nested}), PGP presence shows an initial odds ratio of 2.13 (p < .01) in Model 1, which attenuates to 1.84 (p < .05) in the full model. Each additional category increases odds by 65\% (OR = 1.65, p < .001). For large-vendor status (Table~\ref{tab:logit_large_nested}), the pattern is similar: PGP presence shows an initial OR of 1.94 (p < .01) that attenuates to 1.69 (p < .05), while each additional category increases odds by 169\% (OR = 2.69, p < .001).
\begin{table}[htbp]
\centering
\scriptsize
\setlength\tabcolsep{4pt}
\begin{tabular}{lrrrrr}
\toprule
& \multicolumn{5}{c}{\textbf{DV: Top Vendor (Top 10\%)}}\\
\cmidrule(lr){2-6}
Variable & Model 1 & Model 2 & Model 3 & Model 4 & Model 5\\
\midrule
PGP present             & 2.13**   & 1.83*    & 1.83*    & 1.82*    & 1.84*\\
\# Categories           &          & 1.62***  & 1.62***  & 1.62***  & 1.65***\\
Avg.\ Price (BTC)       &          &          & 1.00     & 1.00     & 1.00\\
Pct.\ Professional      &          &          &          & 0.97     & 0.98\\
Avg.\ Rating            &          &          &          & 2.58*    & 2.61*\\
\addlinespace
\multicolumn{6}{l}{\emph{Primary drug class}}\\
Cannabis                &          &          &          &          & 1.22*\\
Dissociatives           &          &          &          &          & 0.89\\
Ecstasy                 &          &          &          &          & 0.95\\
Opioids                 &          &          &          &          & 0.64***\\
Other Minor             &          &          &          &          & 1.00\\
Prescription            &          &          &          &          & 1.08\\
Psychedelics            &          &          &          &          & 0.79*\\
RCs                     &          &          &          &          & 1.15\\
Steroids                &          &          &          &          & 1.21\\
Stimulants              &          &          &          &          & 0.92\\
\addlinespace
\multicolumn{6}{l}{\emph{Secondary drug class}}\\
Cannabis (sec)          &          &          &          &          & 1.22*\\
Dissociatives (sec)     &          &          &          &          & 0.89\\
Ecstasy (sec)           &          &          &          &          & 0.95\\
Opioids (sec)           &          &          &          &          & 0.64***\\
Other Minor (sec)       &          &          &          &          & 1.00\\
Prescription (sec)      &          &          &          &          & 0.87\\
Psychedelics (sec)      &          &          &          &          & 0.79*\\
RCs (sec)               &          &          &          &          & 1.15\\
Steroids (sec)          &          &          &          &          & 1.21\\
Stimulants (sec)        &          &          &          &          & 0.92\\
\midrule
Pseudo-$R^2$            & 0.008    & 0.112    & 0.112    & 0.118    & 0.134\\
\bottomrule
\multicolumn{6}{l}{\footnotesize $^{*}p<0.05$, $^{**}p<0.01$, $^{***}p<0.001$. $N=2\,653$. Coefficients in odds ratios.}\\
\end{tabular}
\caption{Nested logistic regression models predicting top vendor status (Models 1–5)}
\label{tab:logit_top_nested}
\end{table}

\begin{table}[htbp]
\centering
\scriptsize
\setlength\tabcolsep{4pt}
\begin{tabular}{lrrrrr}
\toprule
& \multicolumn{5}{c}{\textbf{DV: Large Vendor (Above Median)}}\\
\cmidrule(lr){2-6}
Variable & Model 1 & Model 2 & Model 3 & Model 4 & Model 5\\
\midrule
PGP present             & 1.94**   & 1.67*    & 1.68*    & 1.68*    & 1.69*\\
\# Categories           &          & 2.58***  & 2.58***  & 2.58***  & 2.69***\\
Avg.\ Price (BTC)       &          &          & 1.00     & 1.00     & 1.00\\
Pct.\ Professional      &          &          &          & 1.10     & 1.12\\
Avg.\ Rating            &          &          &          & 1.60***  & 1.62***\\
\addlinespace
\multicolumn{6}{l}{\emph{Primary drug class}}\\
Cannabis                &          &          &          &          & 1.00\\
Dissociatives           &          &          &          &          & 0.66***\\
Ecstasy                 &          &          &          &          & 0.82**\\
Opioids                 &          &          &          &          & 0.61***\\
Other Minor             &          &          &          &          & 1.00\\
Prescription            &          &          &          &          & 0.92\\
Psychedelics            &          &          &          &          & 0.77***\\
RCs                     &          &          &          &          & 1.03\\
Steroids                &          &          &          &          & 1.18\\
Stimulants              &          &          &          &          & 0.84**\\
\addlinespace
\multicolumn{6}{l}{\emph{Secondary drug class}}\\
Cannabis (sec)          &          &          &          &          & 1.00\\
Dissociatives (sec)     &          &          &          &          & 0.66***\\
Ecstasy (sec)           &          &          &          &          & 0.82**\\
Opioids (sec)           &          &          &          &          & 0.61***\\
Other Minor (sec)       &          &          &          &          & 1.00\\
Prescription (sec)      &          &          &          &          & 0.85\\
Psychedelics (sec)      &          &          &          &          & 0.77***\\
RCs (sec)               &          &          &          &          & 1.03\\
Steroids (sec)          &          &          &          &          & 1.18\\
Stimulants (sec)        &          &          &          &          & 0.84**\\
\midrule
Pseudo-$R^2$            & 0.006    & 0.168    & 0.168    & 0.172    & 0.185\\
\bottomrule
\multicolumn{6}{l}{\footnotesize $^{*}p<0.05$, $^{**}p<0.01$, $^{***}p<0.001$. $N=2\,653$. Coefficients in odds ratios.}\\
\end{tabular}
\caption{Nested logistic regression models predicting large vendor status (Models 1–5)}
\label{tab:logit_large_nested}
\end{table}

Opioid vendors show substantially reduced odds of achieving both top-vendor (OR = 0.64, $p < .001$) and large-vendor status (OR = 0.61, $p < .001$). Psychedelic vendors also show reduced odds for both outcomes. Cannabis vendors show increased odds of top-vendor status (OR = 1.22, $p < .05$).

\subsection{Predictors of PGP Adoption}

To better understand the relationship between diversification and encryption signaling, we examined factors predicting PGP adoption among vendors. Table~\ref{tab:pgp_predictors} presents a logistic regression with PGP presence as the dependent variable.

\begin{table}[htbp]
\centering
\scriptsize
\setlength\tabcolsep{4pt}
\begin{tabular}{lrrrr}
\toprule
Variable & Coef. & SE & OR & Sig. \\
\midrule
Intercept & -3.892 & 0.451 & 0.02 & *** \\
Number of categories & 0.207 & 0.042 & 1.23 & *** \\
Average rating & 0.074 & 0.089 & 1.08 &  \\
Average price (BTC) & 0.000 & 0.001 & 1.00 &  \\
Total listings & 0.003 & 0.002 & 1.00 &  \\
\midrule
Pseudo-$R^2$ & \multicolumn{4}{c}{0.058} \\
\bottomrule
\multicolumn{5}{l}{\footnotesize $N=2\,653$. $^{*}p<0.05$, $^{**}p<0.01$, $^{***}p<0.001$. Coefficients in odds ratios.}\\
\end{tabular}
\caption{Logistic regression predicting PGP adoption}
\label{tab:pgp_predictors}
\end{table}

The results show that vendors with more categories are significantly more likely to advertise PGP encryption (OR = 1.23, $p < .001$), with each additional category increasing the odds of PGP adoption by 23\%. This suggests that encryption signaling is part of a broader professionalization strategy associated with business diversification rather than an isolated security practice.

\subsection{Exceptional Cases}

Table~\ref{tab:nonpgp_outliers} identifies the five largest vendors who operate without advertising PGP encryption. These vendors demonstrate that PGP signaling is not necessary for achieving substantial market presence, though most still exhibit high diversification across multiple drug categories.

\begin{table}[htbp]
\centering
\scriptsize
\begin{tabular}{lrr}
\toprule
Vendor & Total listings & Categories\\
\midrule
mssource   & 737 & 20\\
RXChemist  & 697 &  7\\
medibuds   & 600 &  3\\
rc4me      & 495 & 13\\
Gotmilk    & 478 & 14\\
\bottomrule
\end{tabular}
\caption{Largest non-PGP vendors}
\label{tab:nonpgp_outliers}
\end{table}

\begin{figure}[htbp]
\centering
\includegraphics[width=.75\textwidth]{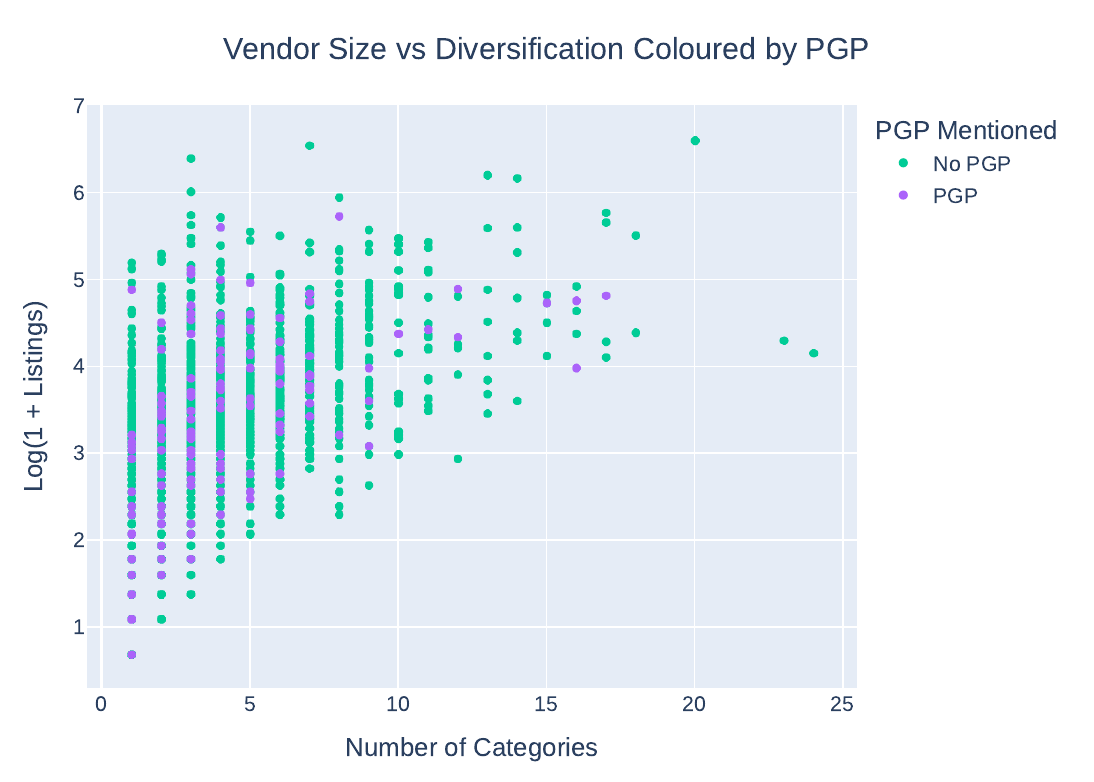}
\caption{Total listings versus diversification, coloured by PGP mention}
\label{fig:scatter_div_size}
\end{figure}

Figure~\ref{fig:scatter_div_size} shows the relationship between total listings and diversification, with points colored by PGP status, confirming the nearly linear relationship between category breadth and listing volume.

\begin{figure}[htbp]
\centering
\includegraphics[width=.7\textwidth]{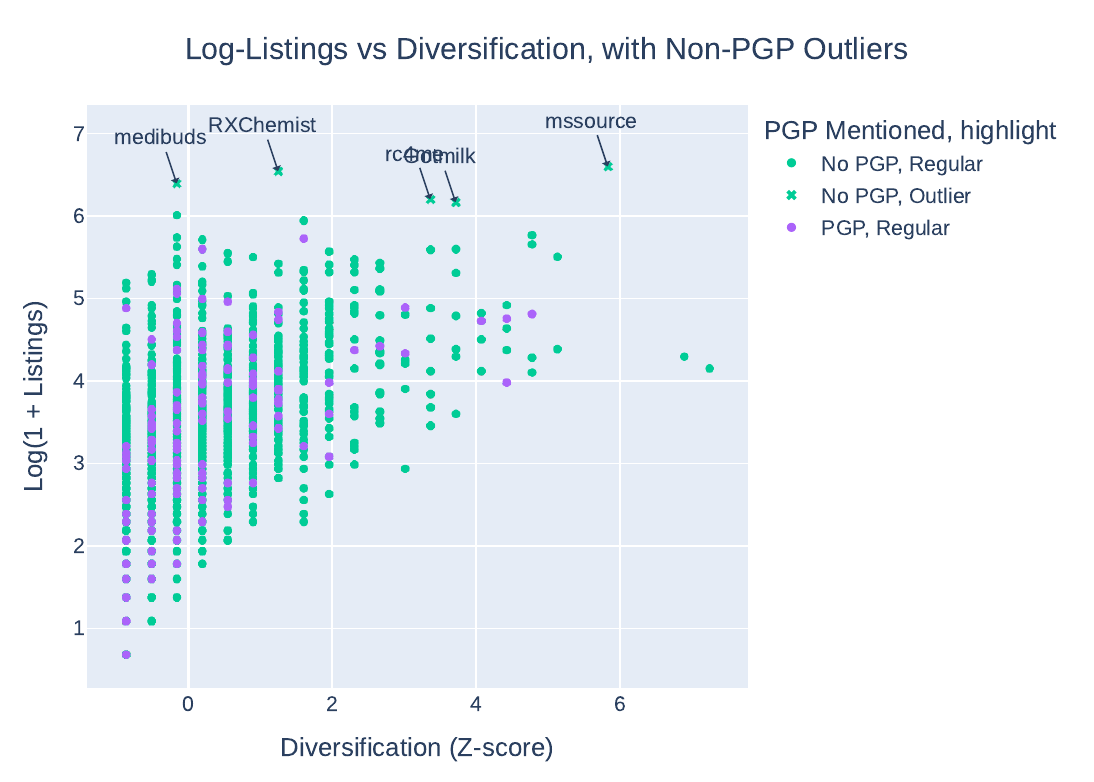}
\caption{Size vs.\ diversification, with non-PGP outliers labelled}
\label{fig:outliers_nonpgp}
\end{figure}

Figure~\ref{fig:outliers_nonpgp} visualizes the exceptional non-PGP vendors within the broader size-diversification relationship, showing they cluster along the high end of both dimensions.

\section{Discussion}

\subsection{Theoretical Contributions}

\subsubsection{Signaling Theory in Anonymous Markets}
Our findings advance signaling theory by demonstrating how trust formation operates in anonymous digital environments where traditional credibility markers are unavailable. The evidence that PGP encryption mentions function as professional signals rather than independent success factors extends signaling theory to illicit commerce contexts, revealing how vendors construct legitimacy through observable behavioral cues when legal institutions and formal business credentials are absent. This finding challenges assumptions about direct utility from cybersecurity practices, suggesting that signaling value may dominate functional value in reputation-dependent markets.

\subsubsection{Diversification Theory in Illicit Commerce}
The dominance of product diversification as a success predictor fundamentally challenges specialization theories traditionally applied to criminal enterprises~\cite{reuter1983disorganized, levitt2000economic}. Our results demonstrate that successful darknet vendors operate as diversified enterprises rather than specialized dealers, revealing a shift in the nature of valued expertise. While traditional dealers benefit from specialized knowledge in production or local distribution~\cite{decker1994legwork}, darknet vendors succeed through logistics and digital marketing skills. This transformation reflects how customer acquisition differs online: street dealers cultivate personal relationships aided by specialization, while darknet customers discover vendors through search functions, incentivizing diversification~\cite{jacques2013code}.

\subsubsection{Digital Market Structure Theory}
The identification of category-specific constraints and systematic scaling differences across drug types contributes to understanding of how market structure varies in digital environments. The finding that opioid vendors face 37\% lower scaling odds compared to other categories suggests potential effects of differential enforcement intensity. During our study period, federal agencies were increasingly targeting opioid distribution~\cite{dea2015nflis}, patterns that intensified dramatically by 2024~\cite{ussc2024annual}. However, without direct enforcement data, alternative explanations including supply constraints or customer preferences cannot be ruled out, highlighting the need for future research linking enforcement actions to vendor outcomes.

\subsection{Managerial Implications}

\subsubsection{Enforcement Strategy Implications}
Our findings suggest that traditional enforcement approaches targeting specific drug categories may prove less effective against diversified vendor operations than previously assumed. The dominance of diversification as a success factor implies that successful vendors can rapidly pivot between product categories when enforcement pressure increases, requiring coordinated multi-category approaches rather than substance-specific targeting. Law enforcement agencies should consider developing capabilities for simultaneous multi-category investigations and monitoring vendor portfolio evolution rather than focusing resources on single drug types.

\subsubsection{Market Intervention Strategies}
The identification of diversification as the primary scaling mechanism suggests that interventions disrupting vendor ability to maintain multiple product categories may prove more effective than traditional supply-side approaches. Strategies that increase coordination costs across categories or create technical barriers to portfolio management could prove more disruptive than enforcement efforts targeting individual substance types.

\subsubsection{Harm Reduction Considerations}
The finding that larger vendors demonstrate higher professional signaling rates suggests potential benefits from policies that encourage vendor professionalization rather than purely punitive approaches. If diversified vendors provide more reliable service and higher product quality, enforcement strategies that drive market fragmentation toward smaller, less professional operators might inadvertently increase health risks for drug users through reduced quality control and service reliability.

\subsection{Policy and Practice Integration}
These findings have important implications for both enforcement strategy and digital marketplace regulation more broadly. The resilience provided by diversification strategies suggests that darknet markets may be more robust to traditional enforcement methods than previously understood, requiring fundamental reconsideration of intervention approaches. The emergence of sophisticated signaling mechanisms and professional standards within illicit markets demonstrates the adaptability of commercial institutions even in environments without formal legal frameworks. These insights underscore the importance of evidence-based approaches that recognize how digital transformation fundamentally alters the structure and dynamics of illicit commerce.

\bibliographystyle{plain}
\section*{Appendix A: Complete Professionalism Term List}

The following terms were used to identify professional practices in vendor listings using case-insensitive string matching:

blister pack, business days, discreet shipping, erowid, escrow, guarantee, lab tested, lab-tested, pharmaceutical name, public key, purity, read profile, refund, reship, sealed, tracking, tracking number, vacuum sealed, vacuum-sealed.

Terms not included in the professionalism classifier: ``stealth'' and ``discreet'' (too general), ``tablet,'' ``pack,'' and ``bottle'' (common product descriptors rather than professionalism indicators).

\section*{Appendix B: Drug Category Classification Methodology}
\paragraph{Primary Classification} We extract the most commonly used category designation from the main hierarchical category field in the format \texttt{Drugs/[Category]/[Subcategory]}. For example, a listing categorized as \texttt{Drugs/Cannabis/Weed} receives a primary classification of Cannabis.

\paragraph{Secondary Classification} We derive the second most commonly used classification from the second-level hierarchical token. Using the same example, \texttt{Drugs/Cannabis/Weed} yields a secondary classification token of Cannabis.

\paragraph{Category Definitions} Our ten-category taxonomy uses the following search terms:

\begin{table}[ht]
\centering
\scriptsize
\begin{tabular}{|l|p{10cm}|}
\hline
\textbf{Category} & \textbf{Search Terms} \\
\hline
Cannabis & cannabis, marijuana, weed, thc, cbd, hash, hashish, ganja, hemp \\
\hline
Dissociatives & ketamine, pcp, dxm, dextromethorphan, mxe, 3-meo-pcp, diphenidine \\
\hline
Ecstasy & mdma, ecstasy, molly, mda, 6-apb, methylone \\
\hline
Opioids & heroin, oxycodone, fentanyl, morphine, codeine, tramadol, opium, hydrocodone \\
\hline
Prescription & adderall, xanax, valium, ritalin, klonopin, ativan, ambien, viagra \\
\hline
Psychedelics & lsd, psilocybin, mushrooms, dmt, mescaline, 2c-b, 2c-i, 25i-nbome \\
\hline
RCs & research chemical, rc, nbome, 25i, 4-aco, 4-ho, 2c-, novel psychoactive \\
\hline
Steroids & testosterone, anavar, winstrol, deca, trenbolone, hgh, growth hormone \\
\hline
Stimulants & cocaine, amphetamine, methamphetamine, speed, crystal, crack \\
\hline
OtherMinor & tobacco, nicotine, kratom, salvia, spice, synthetic marijuana \\
\hline
\end{tabular}
\caption{Drug Category Classification Terms}
\label{tab:drug_terms}
\end{table}

\bibliography{references}

\end{document}